\documentclass[twocolumn,prb,floatfix,unsortedaddress]{revtex4}
\usepackage{graphicx, subfigure, amsfonts,amssymb,amsmath, array, gensymb,fontenc,color}

\begin{document}

 % TITLE AND ABSTRACT
 
\author{Michele Pizzochero}
\email{michele.pizzochero@epfl.ch} 
\author{Oleg V. Yazyev}
\email{oleg.yazyev@epfl.ch} 
\affiliation{Institute of Physics, Ecole Polytechnique F\'ed\'erale de Lausanne (EPFL), CH-1015 Lausanne, Switzerland}
\date{\today}
\title{Point defects in the 1\emph{T'} and 2\emph{H} phases of single-layer MoS$_2$: \\
A comparative first-principles study}%

\begin{abstract}
The metastable \emph{1T'} phase of layered transition metal dichalcogenides has recently attracted considerable interest due to electronic properties, possible topological electronic phases and catalytic activity.
We report a comprehensive theoretical investigation of intrinsic point defects in the \emph{1T'} crystalline phase of single-layer molybdenum disulfide (\emph{1T'}-MoS$_2$), and provide comparison to the well-studied semiconducting \emph{2H} phase. Based on density functional theory calculations, we explore a large number of configurations of vacancy, adatom and antisite defects and analyse their atomic structure, thermodynamic stability, electronic and magnetic properties. 
The emerging picture suggests that, under thermodynamic equilibrium, \emph{1T'}-MoS$_2$  is more prone to hosting lattice imperfections than the \emph{2H} phase. More specifically, our findings reveal that the S atoms that are closer to the Mo atomic plane are the most reactive sites. Similarly to the \emph{2H} phase, S vacancies and adatoms in \emph{1T'}-MoS$_2$ are very likely to occur while Mo adatoms and antisites  induce local magnetic moments. Contrary to the \emph{2H} phase, Mo vacancies in \emph{1T'}-MoS$_2$  are expected to be an abundant defect due to the structural relaxation that plays a major role in lowering the defect formation energy.
Overall, our study predicts that the realization of high-quality flakes of \emph{1T'}-MoS$_2$  should be carried out under very careful laboratory conditions but at the same time the facile defects introduction can be exploited to tailor physical and chemical properties of this polymorph.
\end{abstract}

\pacs{} 
\maketitle

% INTRODUCTION

\section{Introduction}

Since the very first isolation of graphene\cite{nov04}, a constantly growing effort has been devoted to the investigation of atomically-thin crystals\cite{nov05}. Within this increasingly large family of materials, single-layer transition metal dichalcogenides\cite{chh13} (TMDs) are expected to play a prominent role in the next-generation technologies. Such an expectation is based on the recent realization of TMD field effect transistors for flexible low-power electronics\cite{rad11,mwchen17}, circuits capable of performing digital logic operation\cite{rad11bis}, gas sensors\cite{li12}, transparent devices for optoelectronics\cite{wan12}, and many more.

Transition metal dichalcogenides are inorganic materials of general formula MX$_2$, with M being a transition metal atom and X a chalcogen atom (S, Se or Te)\cite{chh13,manz17}. The most representative member of this novel class of systems is undoubtedly MoS$_2$. Similarly to graphene, MoS$_2$ can be obtained through micromechanical or chemical exfoliation from molybdenite, a largely available van der Waals layered mineral well-known in the field of lubricant industry, as well as via chemical vapor deposition or chemical vapor transport techniques\cite{yaz15}.
However, contrary to graphene, MoS$_2$ (and, in general, all the layered TMDs) admits several polymorphs differing in the coordination of the Mo atom\cite{cal13}, \emph{i.e.} the \emph{2H},  \emph{1T} and  \emph{1T'} phases, which exhibit different structural and electronic properties, depending on the stacking of the S/Mo/S atomic planes. In the case of MoS$_2$, the most stable phase is the \emph{2H} phase, where the three layers are stacked in an \textsc{aba} order and a direct band gap is observed in the electronic spectrum. Conversely, the metallic \emph{1T} phase is characterized by the \textsc{abc} stacking order. This \emph{1T} phase, however, is not stable with respect to the dimerization distortion of Mo atoms giving rise to the \emph{1T'} phase\cite{hei99,cal13}, which is the main focus of this work.

The interest in the metastable \emph{1T'} phase of MoS$_2$, as well as of other group VI TMDs,  is mainly due to its peculiar electronic properties, where the combination of band inversion and a small band gap induced by spin-orbit interactions was predicted to result in the topological Quantum Spin Hall (QSH) insulator phase. The band gap magnitude of the QSH phase in \emph{1T'} TMDs can be controlled by an external electric field as well as lattice strain\cite{qian14,pul16}. Based on these novel electronic properties, a \emph{1T'}-MoS$_2$  field effect transistor has been proposed\cite{qian14,liulei15}. Additionally, \emph{1T'} and \emph{2H} phases can coexist in the same monolayer, thereby yielding to the formation of a metal-semiconductor lateral heterojunction that can be exploited for the realization of novel nanodevices\cite{eda12}. It has been shown that \emph{1T'}-TMDs have superior catalytic activity in hydrogen evolution reaction compared to the \emph{2H} phase\cite{guogao15,seok17} and their peculiar geometries drive the accommodation of organic cations towards specific lattice sites, with potential applications in TMDs-based sensors\cite{golo15}. Several strategies have been proposed in order to promote a transition between the two crystalline phases\cite{voi15} including lattice strain\cite{due14}, lithiation\cite{esf15}, mild annealing\cite{eda11}, alloying\cite{raf16} as well as electrostatic gating\cite{li16}$- $ with the aim of exploiting different functionalities within the same material.

The presence of certain amount of defects in crystalline materials is inevitable, and in this respect TMDs are no exception. On one hand, crystalline disorder can be detrimental to various material's properties. On the other hand, defects can be deliberately introduced offering novel opportunities for tailoring properties of the material. Defects in TMDs can be created in different ways, e.g. by means of electron beam irradiation\cite{kom12, kom13}, argon plasma treatment\cite{cho15}, thermal annealing\cite{don13} and $\alpha$-particle or ion bombardment\cite{mat12}.
Defects in the \emph{2H} phase of MoS$_2$ have been extensively studied both theoretically\cite{kom15, noh14, kc14, gon16} and experimentally\cite{hon15,zho13, liu16,naj15}. Beside the fundamental interest, introduction of crystalline lattice imperfections was used to tailor technologically relevant properties of this material\cite{lin16}. Just to mention a very few examples, sulfur vacancies were shown to induce localized states\cite{pan16} that enhance optical absorption\cite{shu16}, lead to electronic transport \emph{via} hopping in the low carrier density regime\cite{qiu13}, promote catalytic activity\cite{li16bis} and degrade thermal conductivity\cite{wan16}. Under strain both Mo and S vacancies are expected to induce magnetism\cite{tao14} and similar predictions were also reported for metal adatoms and antisite defects\cite{ata11bis,Zhe15,cao16}, whereas a proper vacancy functionalization can lead to giant magnetocrystalline anisostropy\cite{siv16}. Magnetic ordering was also experimentally observed in MoS$_2$ at room temperature upon proton irradiation, and its origin attributed to the defect formation\cite{siv16}. Additionally, both vacancies and adatoms were used to tailor the electric properties of MoS$_2$ by engineering the band gap width\cite{kva15, wan16bis}.

So far, this large body of works focusing on various crystalline lattice imperfections was restricted to the stable \emph{2H} phase only,  while the impact of intrinsic point defects on the properties of the \emph{1T'}-phase TMDs has not been covered yet, although samples of this metastable polymorph have been widely characterized by microscopy techniques\cite{wan16}. It is therefore important to address this issue theoretically in order to provide insights into experiments and portray a wide-angle view of the effect of intrinsic atomic-scale defects on the properties of \emph{1T'}-MoS$_2$. In this paper, we report on the atomic structure and thermodynamic stability of vacancy, antisite and adatom defects in single-layer \emph{1T'}-MoS$_2$ investigated by means of first-principles calculations. Throughout this work, we systematically compare defects in the \emph{1T'}-phase of monolayer MoS$_2$ with those in the \emph{2H} polymorph. 
Even though we focus on MoS$_2$, the results of our study can be reasonably extended to other group-VI disulfides and diselenides (\emph{i.e.}  MoSe$_2$, WS$_2$ and  WSe$_2$), where a similar physics of defects is expected.

This paper is organized as follows: Section II presents the details of our calculations. In Section III we discuss the results of our simulations. Finally, Section IV summarizes and concludes our work.

%METHODS

\section{Methodology}
\subsection{First-principles calculations}

Our first-principles calculations have been performed within the density functional theory (DFT) formalism\cite{jon15} as implemented in the \textsc{siesta} code\cite{sol02}. The exchange and correlation effects are treated within the semilocal density functional of Perdew, Burke and Erznerhof\cite{per96,per97}. Core electrons were replaced by norm-conserving pseudopotentials\cite{ham79} generated within the Troullier-Martins approach\cite{tro91} whereas Kohn-Sham wavefunctions for valence electrons were expanded in a linear combination of double-$\zeta$ plus polarization (DZP) basis set in conjunction with a mesh cutoff of 250 Ry. Throughout this work, we neglect spin-orbit interaction as its contribution to the formation energies of defects is negligible, but we include spin-polarization in order to unravel possible defect-induced magnetism and correctly treat its larger contribution to the formation energies.
However, spin-orbit interaction was considered when addressing the magnetic anisotropy of defects that give rise local magnetic moments. We performed calculations of total energies by constraining the resulting magnetic moment in the in-plane direction ($E_{\parallel}$) and in the out-of-plane direction ($E_{\perp}$). The magnetocrystalline anisotropy energy (MAE) is calculated as the energy difference between the two configurations ($E_{\rm MAE} = E_{\parallel} - E_{\perp}$) and assumes negative (positive) values when the magnetic moment has lower energy in the case of in-plane (out-of-plane) orientation.

In order to model isolated point defects we considered rectangular 3 $\times$ 6 supercells of monolayer MoS$_2$ containing 108 atoms. This corresponds to supercell dimensions of 17.24 {\AA} $\times$ 19.15 {\AA} for the \emph{1T'} phase and 16.61 {\AA} $\times$ 19.21 {\AA} for the \emph{2H} phase. The periodic replicas were separated by 16 {\AA} in the out-of-plane direction. Integration over the first Brillouin zone was performed with a $\Gamma$-centered grid following the scheme devised by Monkhorst and Pack (MP)\cite{mon76}. Specifically, we used the equivalent of 12 $\times$ 18 $\times$ 1 $k$-points for geometry optimization and a five time denser mesh of 60 $\times$ 90 $\times$ 1 $k$-points per rectangular 6-atoms unit cell for the calculation of the electronic density of states (DOS), where a broadening of 0.03 eV was employed. The number of $k$-points was properly reduced for the above mentioned supercell in order to preserve the MP grid density. Geometries were considered relaxed when the maximum force component acting on each atom is lower than 0.01 eV/{\AA}.  During the structural optimization of the defect models supercell dimensions were kept fixed to the values of the pristine system, according to the recommendation of Ref. \onlinecite{wal04}. We systematically included the slab-dipole correction\cite{ben99} in order to eliminate the effect of possible artificial electric fields.

\subsection{Formation energies}
Formation energy of defects is the primary property that allows understanding their relative stability under thermodynamic equilibrium. Since MoS$_2$ is a binary system, in most cases point defects result in the deviation from nominal stoichiometry, which makes their formation energies dependent on chemical potentials of the constituent elements. The formation energy $E_{form}(\mu)$ is defined as
\begin{equation}
E_{form}(\mu) = E_{def} - E_{clean} - \Delta N \mu ,
\label{formationenergy}
\end{equation}
with $E_{def}$ and $E_{clean}$ being the total energies of the defect model and pristine materials, respectively, while $\Delta N$ is the change of the number of atoms upon introducing the defect and $\mu$ the chemical potential of one of the chemical elements. The calculation of the formation energy requires the choice of the reference elemental system for the determination of the chemical potential. In our work, as a reference system we choose bulk sulfur. The corresponding total energy of bulk sulfur is obtained by relaxing the experimental crystal structure of the $\alpha$ phase\cite{ret87}, which consists of S$_8$ puckered rings packed in a orthorhombic lattice, with a intramolecular S--S bond length of 2.13 \AA.

The boundaries of the relevant range of the chemical potential of sulfur $\mu_S$ are defined by the conditions at which precipitation of one of the chemical elements takes place. The upper boundary, corresponding to S-rich condition, is given by $\mu^{max}_S = E_S$ with $E_S$ being the energy per atom in the $\alpha$ phase of bulk sulfur. The lower boundary that corresponds to Mo-rich conditions is defined as $\mu^{min}_S = (E_{clean} - E_{Mo})/2$, where $E_{\mathrm Mo}$ is the total energy per atom in the bcc crystal structure of bulk molybdenum. Specifically, the ranges of stability of the monolayers are 0 $< $ $\mu_S$ $< $ $-$1.25 eV for \emph{1T'}-MoS$_2$ and 0 $< $ $\mu_S$ $< $ $-$1.54 eV for \emph{2H}-MoS$_2$.  
Such a difference in the boundary of $\mu_S$ implies the existence of a chemical potential window of 0.29 eV in the Mo-rich conditions, where the most stable phase is preferred against bulk Mo precipitation.
%RESULTS AND DISCUSSION%

\section{Results and discussion}

%PRISTINE MOS2%

\subsection{Pristine MoS$_2$}
\begin{figure}[]
  \centering
 \includegraphics[width=1\columnwidth]{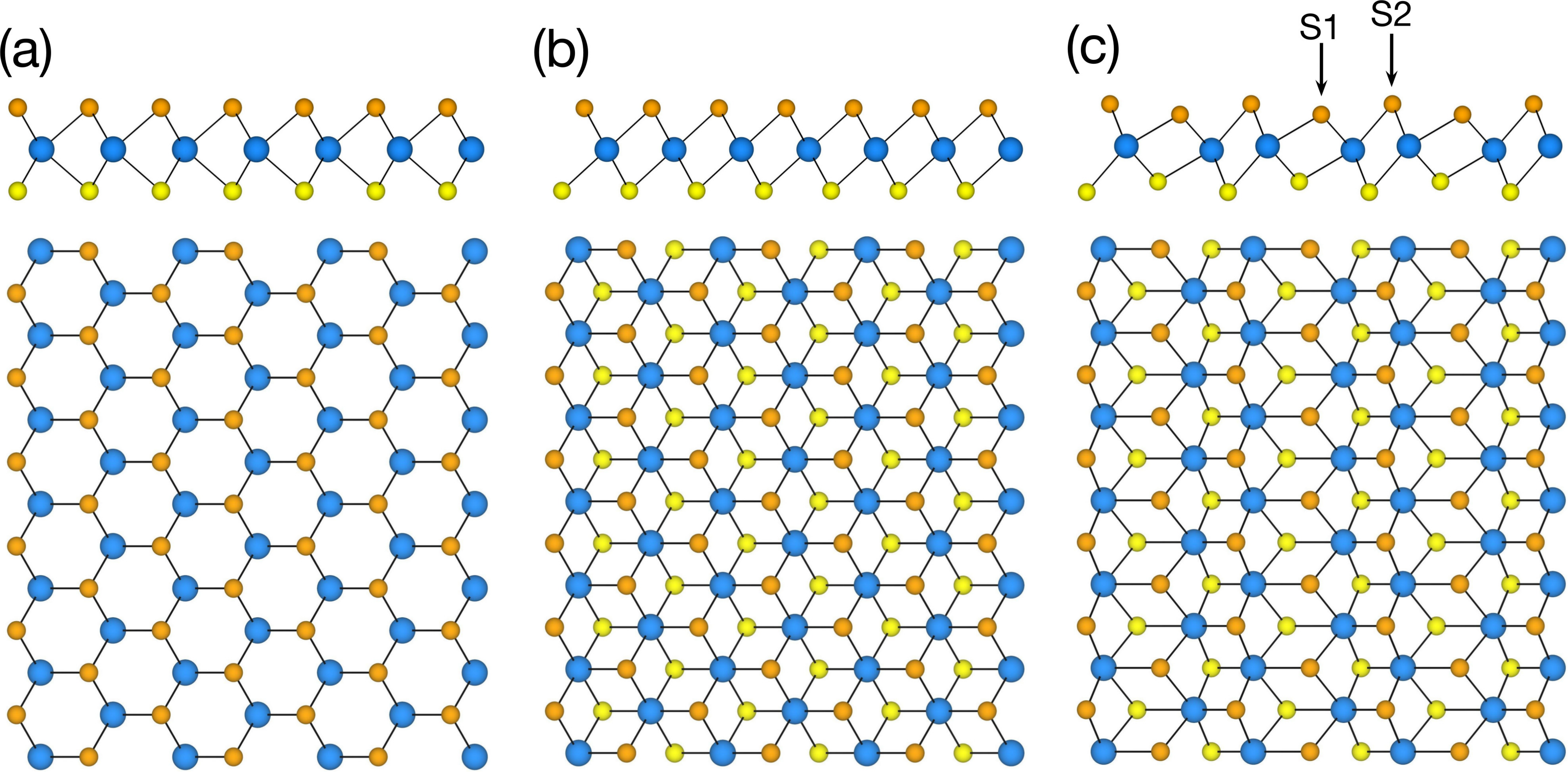}
  \caption{Side (upper panels) and top (lower panels) views of the atomic structures of (a) \emph{2H}, (b) \emph{1T} and (c) \emph{1T'} crystalline phases of monolayer MoS$_2$. Blue balls represent Mo atoms, whereas orange (yellow) balls represent S atoms belonging to the top (bottom) atomic planes.  \label{Geom-Pristine}}
\end{figure}

%\begin{table}
%\caption{\label{Tab-1}Formation energies ($E_{form}^{ideal}$) and interatomic distances (d$\textsubscript{Mo--S}$) of the ideal crystalline phases of single-layer MoS$_2$.}
%\begin{ruledtabular}
%\begin{tabular}{llll}
% & \emph{2H}  & \emph{1T}  & \emph{1T'} \\
% \hline\\
%$E_{form}^{ideal}$ (eV/f.u.)  &$-$3.07  &$-$2.25  & $-$2.50\\
%d$\textsubscript{Mo--S}$ ({\AA})  & 2.44  & 2.46  & 2.50, 2.42\\
%\end{tabular}
%\end{ruledtabular}
%\end{table}

\begin{figure*}
  \centering
  \includegraphics[width=1.8\columnwidth]{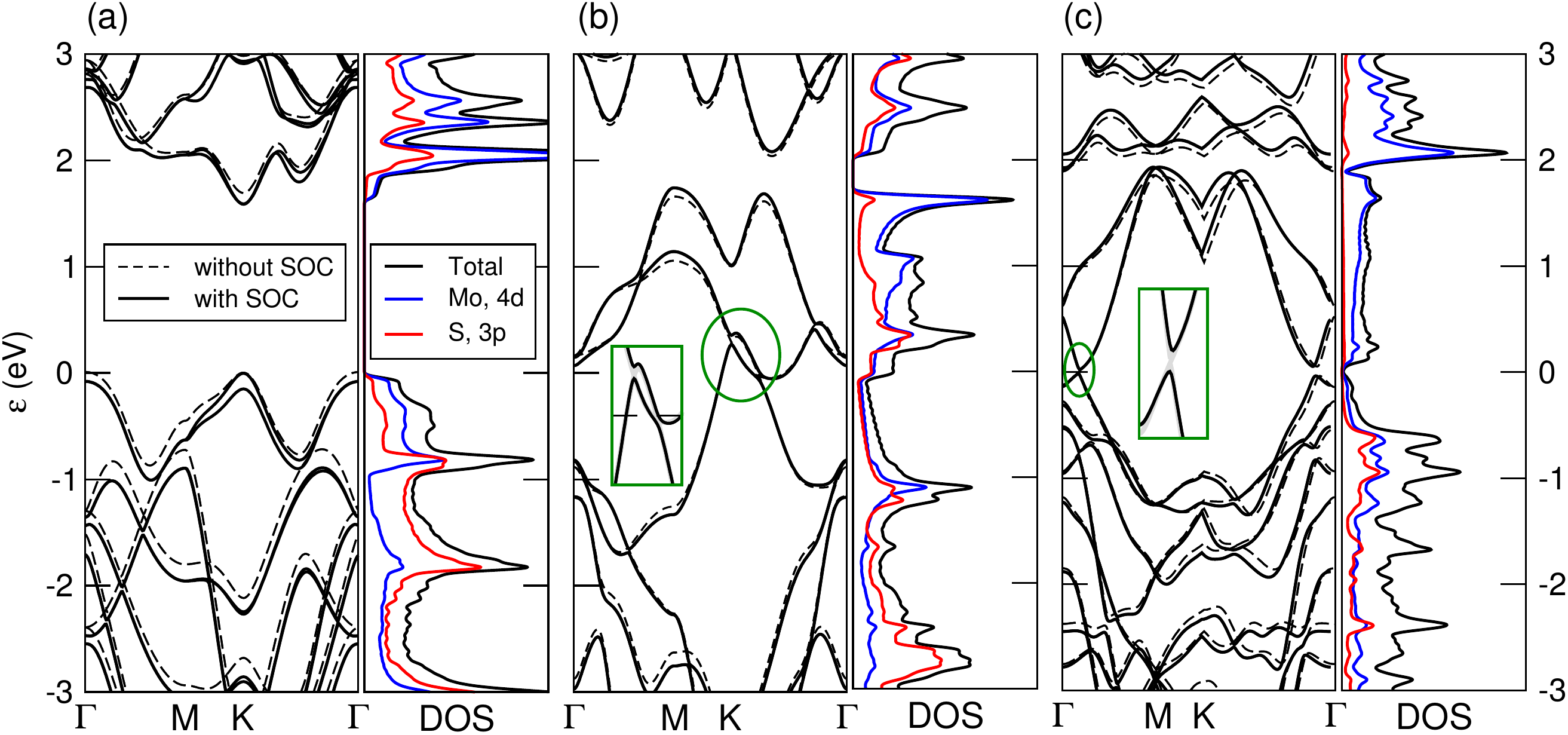}
  \caption{Electronic band structures with and without spin-orbit coupling (SOC) (left) and  density of states  (DOS) plots (right) for the (a) \emph{2H}, (b) \emph{1T} and (c) \emph{1T'} crystalline phases of single-layer MoS$_2$.  Band structures have been obtained using a hexagonal 1 $\times$ 1 supercell for \emph{2H}- and \emph{1T}-MoS$_2$ and a 2 $\times$ 1 supercell for \emph{1T'}-MoS$_2$ along high-symmetry points $\Gamma$(0;0), $M$($\frac{1}{2}$;0) and $K$($\frac{1}{3}$;$\frac{1}{3}$).  Green circles in (b) and (c) indicate the band crossing at the Fermi level and the insets are their magnified views. Valence band maximum of the \emph{2H}- and Fermi levels in \emph{1T}- and \emph{1T'}-MoS$_2$ band structures are set to zero.\label{DOS-Pristine}}
\end{figure*}

Before presenting the results for defect models,  for the sake of completeness we overview our calculations of pristine MoS$_{2}$ monolayers in various crystalline phases, whose atomic structures are shown in Fig. \ref{Geom-Pristine}.

Single-layer MoS$_{2}$ consists of a Mo atoms plane sandwiched between two atomic planes of S atoms, and the stable polymorph is the \emph{2H} phase (Fig. \ref{Geom-Pristine}a), where the S/Mo/S atomic planes are arranged to form an \textsc{aba} stacking order, with a Mo--S interatomic distance of 2.44 {\AA} and a trigonal prismatic coordination of Mo atoms. As shown in Fig. \ref{DOS-Pristine}a, this phase is semiconducting with a direct band gap of 1.7 eV  and band edges mostly composed of Mo \emph{4d} atomic orbitals and S \emph{3p} atomic orbitals.

A rhombohedral \textsc{abc} stacking order of the S/Mo/S atomic planes corresponds to the \emph{1T} phase, where the Mo atom has a distorted octahedral coordination (Fig. \ref{Geom-Pristine}b). In agreement with previous calculations\cite{eny11}, this structure is found to be 0.82 eV/f.u. less stable than the \emph{2H} phase and presents a longer Mo--S interatomic distance of 2.46 {\AA}. The different stacking order has deep consequences on the electronic structure around the Fermi level, since this phase shows a metallic character, as shown in Fig. \ref{DOS-Pristine}b. As pointed out by several works, this system is dynamically unstable in its free-standing form, and large imaginary branches throughout the first Brillouin zone appear in its phonon dispersion\cite{shi14, cal13}.

\emph{1T}-MoS$_2$ undergoes a dimerization distortion resulting in the so-called \emph{1T'} phase (Fig. \ref{Geom-Pristine}c). Though 0.57 eV/f.u. less stable than \emph{2H}-MoS$_2$, the metastable \emph{1T'} phase is a local minimum on the Born-Oppenheimer surface, as supported by the absence of imaginary frequencies in the phonon spectrum reported by several authors \cite{fan14, cal13}. Contrary to the other phases, from Fig. \ref{Geom-Pristine}c one can see that in the \emph{1T'} phase there exist two inequivalent sulfur atoms, whose Mo--S interatomic distances are 2.50~{\AA} and 2.42~{\AA}.
To help distinguishing these S atoms, throughout this work we label S1 (S2) the sulfur atom closer to (farther from) the Mo atomic plane, as shown in Fig. \ref{Geom-Pristine}c. 
The longer Mo-S1 bond, as opposed to Mo-S2 bond length, clearly signals an inhomogeneity in the bond strength, thereby suggesting that the bond involving the S atom closer to the Mo plane (the S1 lattice site) is weaker than the other one.
Contrary to the \emph{2H} phase, single-layer \emph{1T'}-MoS$_2$ has a semimetallic band structure where valence and conduction bands degeneracy at the Fermi level is lifted by the spin-orbit interaction (Fig. \ref{DOS-Pristine}c).

Despites the considerable differences in the properties of the three polymorphs, one can still recognize common features in their DOS plots (Fig. \ref{DOS-Pristine}). For all the phases we observe a sharp peak due to the Mo \emph{d} atomic orbitals at $\approx$2~eV as well as a peak corresponding to the \emph{p} and \emph{d} atomic orbitals of S and Mo, respectively, at about $-$1~eV. 
Comparing the electronic structure of the \emph{2H} phase with the \emph{1T'} phase, some strong analogies in the shape and composition in the energy range between $-$1~eV and 2~eV are apparent; the main difference is due to the population of the band-gap region of \emph{2H}-MoS$_2$ by the \emph{4d} states of Mo in the \emph{1T'} polymorph, which confers a semimetallic nature to the latter phase. 

Throughout the rest of the paper, we analyze the structure, energetics and electronic properties of point defects,  \emph{i.e.}  vacancy, adatom and antisite defects, in the crystalline phases of single-layer MoS$_2$. We restrict our further investigations to the \emph{2H} and \emph{1T'} polymorphs and systematically discuss their similarities and differences.

%VACANCIES%

\subsection{Vacancy Defects}
\begin{figure}[b]
  \centering
  \includegraphics[width=1\columnwidth]{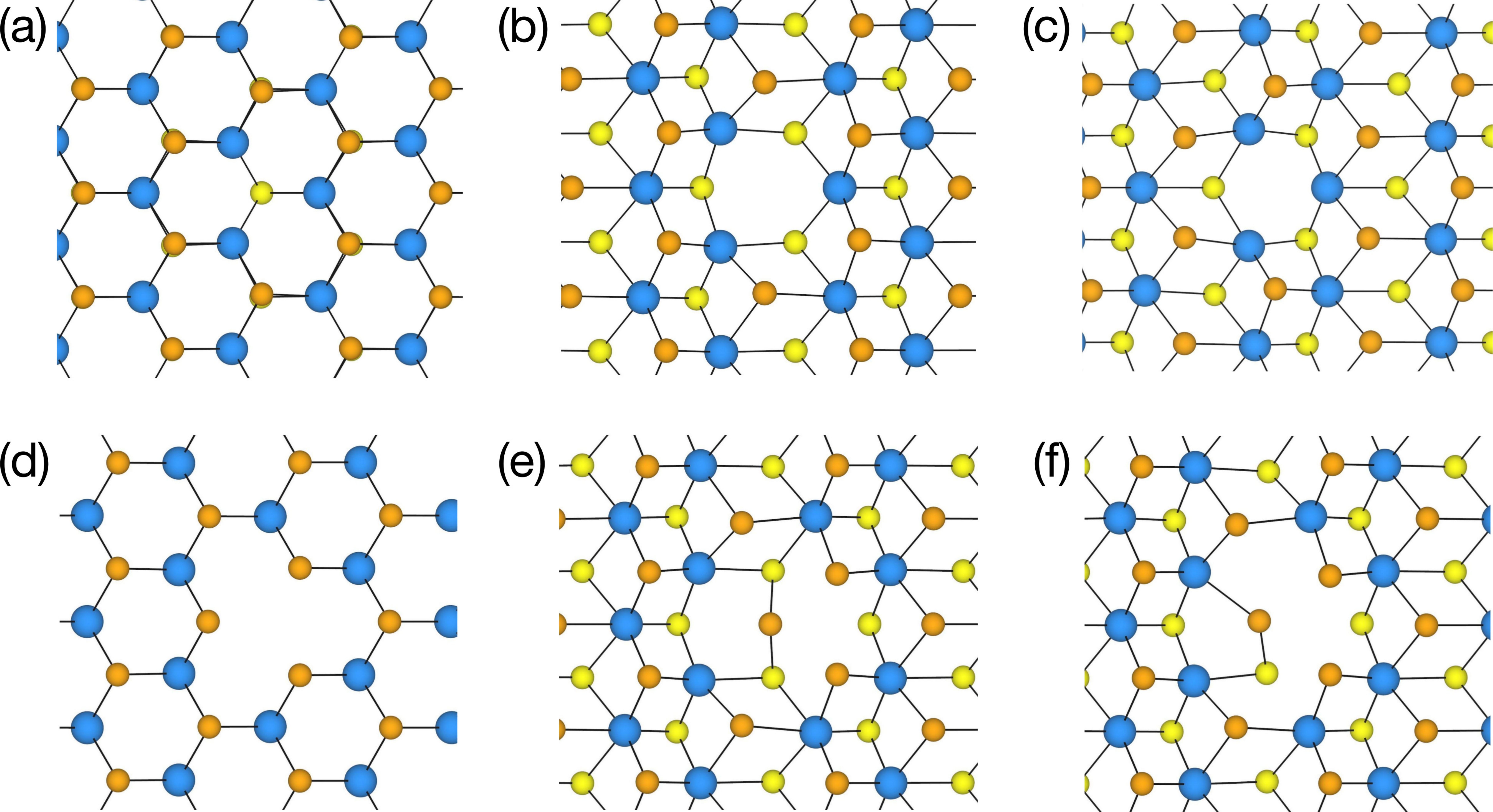}
  \caption{Atomic structures of single-atom vacancy defects: (a)  \emph{2H}-V$\textsubscript{S}$ (b) \emph{1T'}-V$\textsubscript{S1}$, (c) \emph{1T'}-V$\textsubscript{S2}$ (d) \emph{2H}-V$\textsubscript{Mo}$, (d) \emph{1T'}-V$\textsubscript{Mo}$ and (e) \emph{1T'}-V$\textsubscript{MoS}$. \label{VMoS}}
\end{figure}

\begin{figure}[]
  \centering
  \includegraphics[width=1\columnwidth]{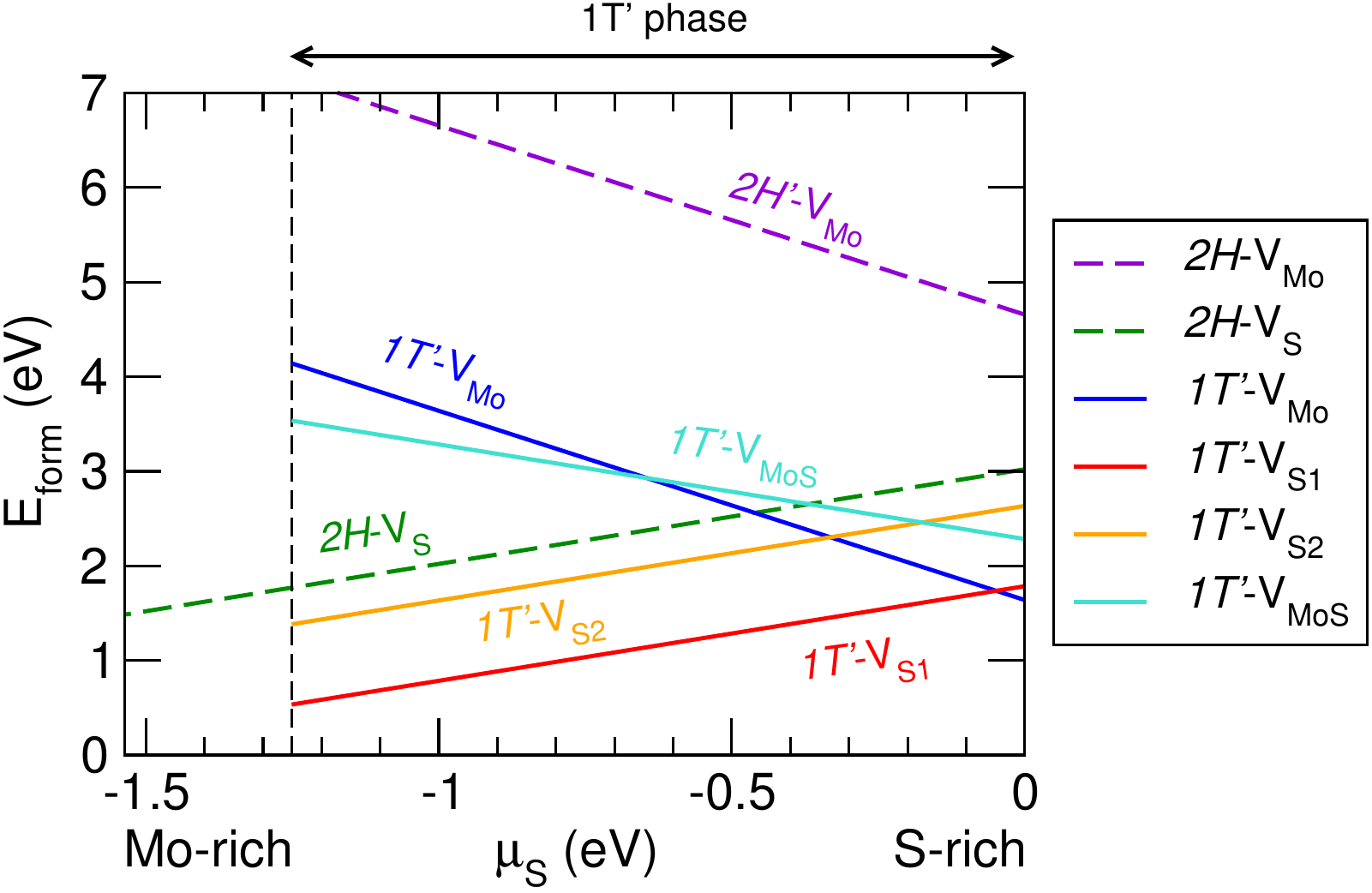}
  \caption{Formation energies of vacancy defects as a function of chemical potential of sulfur $\mu_S$. Vertical dashed line indicates the range of stability of \emph{1T'}-MoS$_{2}$. Colored solid (dashed) lines represent the formation energies of defects in the \emph{1T'} (\emph{2H}) phase. See text for details on the chemical potential ranges. \label{FE-Vacancies}}
  \end{figure}

First, we consider single vacancies (Figs. \ref{VMoS}a-d). For both \emph{1T'} and \emph{2H} polymorphs, the formation energies as a function of chemical potential of sulfur $\mu_S$ are presented in Fig. \ref{FE-Vacancies}. As expected, formation of a S (Mo) vacancy is more likely in a Mo-rich (S-rich) environment.

Upon introducing a S vacancy in the \emph{2H} phase, the Mo--Mo interatomic distance between the three metal atoms in the vicinity of the missing atom is reduced from 3.20~{ \AA} to 3.09~{\AA}. This latter value is in excellent agreement with experiments, where aberration-corrected transmission electron microscopy (AC-TEM) revealed such a distance to be equal to 3.10 {\AA} \cite{wan16bis}. In the \emph{1T'} polymorph, because of the presence of two inequivalent sulfur  atoms S1 and S2, two distinct S vacancy configurations, \emph{1T'}-V$\textsubscript{S1}$ and \emph{1T'}-V\textsubscript{S2}, respectively, are possible. In both configurations, the formation of a single vacancy affects the average distance between Mo atoms surrounding the defective site. Specifically, for \emph{1T'}-V$\textsubscript{S1}$ (\emph{1T'}-{V}$\textsubscript{S2}$) these average distances increase from 3.61~{\AA} (2.92~{\AA}) in the pristine crystal to 3.78~{\AA} (3.20~{\AA}) upon vacancy formation. Furthermore, the inequivalence of the two S atoms is strongly reflected in the relative stabilities of vacancy defects, where the formation energy \emph{1T'}-V$\textsubscript{S1}$ is 0.85~eV lower than that of \emph{1T'}-V$\textsubscript{S2}$ defect. This agrees with the fact that Mo--S1 bond is weaker than the Mo--S2 bond as explained in the previous section.
In general, as reported in Fig. \ref{FE-Vacancies}, single S vacancies form easier  in the \emph{1T'} phase than in the \emph{2H} phase, presumably because of the lower stability of the former polymorph with respect to the latter by 0.57 eV/f.u.

\begin{figure}[]
  \centering
  \includegraphics[width=1\columnwidth]{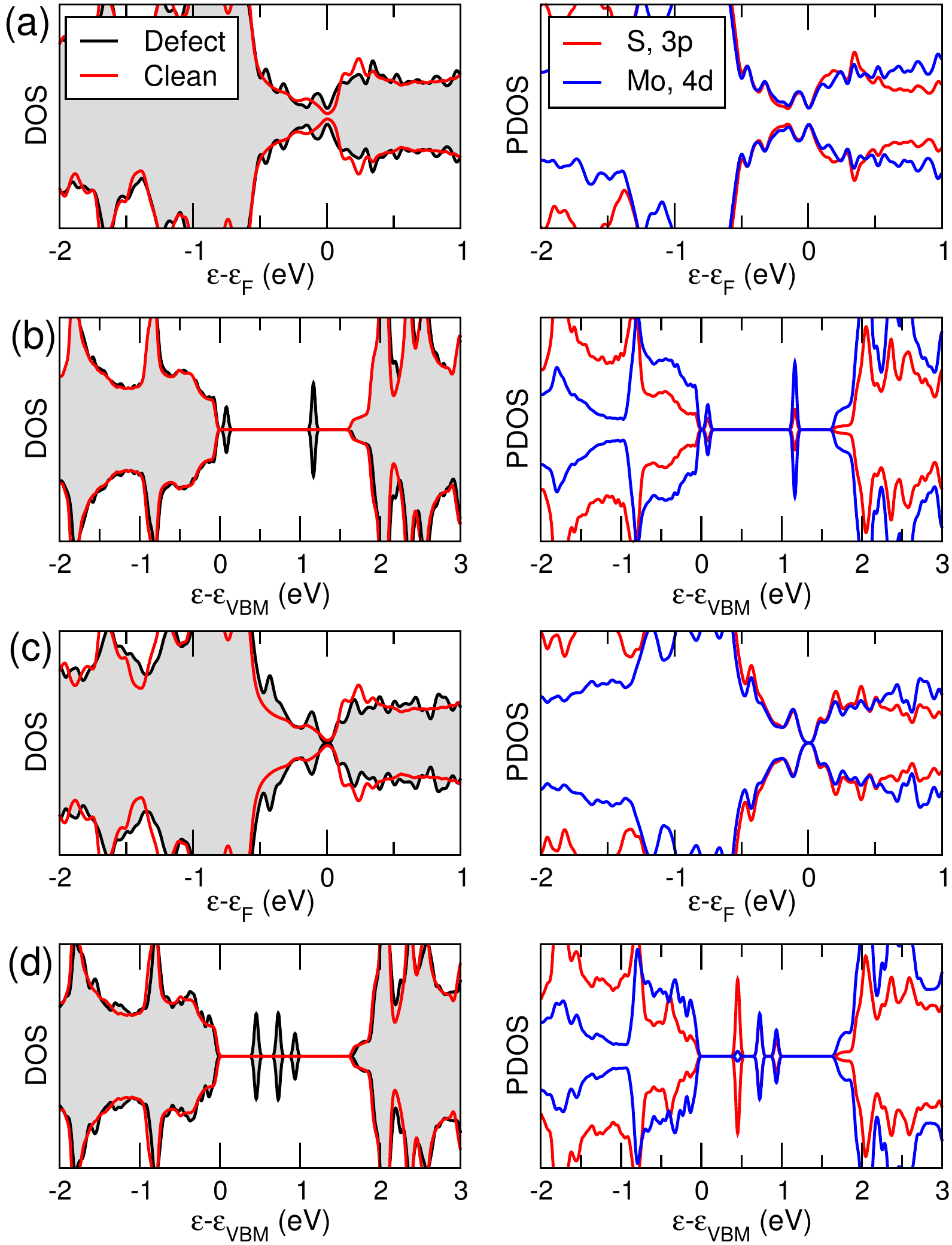}
  \caption{Total (left) and projected (right) electronic density of states plots of (a) \emph{1T'}-V$\textsubscript{S1}$, (b) \emph{2H}-V$\textsubscript{S}$, (c) \emph{1T'}-V$\textsubscript{Mo}$  and (d) \emph{2H}-V$\textsubscript{Mo}$ defects. The total DOS of the pristine (red lines) and the defective (black lines) MoS$_2$ monolayers are shown.  \label{DOS-Vacancies}}
\end{figure}

According to previous reports\cite{pan16}, the chalcogen vacancy in the \emph{2H} phase leads to the formation of two mid-gap states, one located very close to the valence band edge and the other $\approx$1~eV higher in energy, deep in the gap (Fig. \ref{DOS-Vacancies}b). Since a missing sulfur atom affects only the coordination sphere of a Mo atom, these states are mainly due to Mo \emph{4d} atomic orbitals. Defect-induced states in the \emph{1T'} phase are hybridized with the bulk states due to the nearly gapless character of \emph{1T'}-MoS$_2$, and characterized both by S- and Mo- atomic orbitals contributions. However, they can still be distinguished, mostly within 0.5~eV below the Fermi level, as shown in Fig.~\ref{DOS-Vacancies}a.

We now consider Mo vacancy defects. For the \emph{2H} phase, the Mo vacancy defect is characterized by a large formation energy as metal atom removal affects coordination spheres of numerous S atoms. This agrees with experimental reports, where Mo vacancies are only rarely observed\cite{hon15}. A very different situation is found for the \emph{1T'} phase, where in the S-rich conditions of Fig. \ref{FE-Vacancies} the formation energy of V$\textsubscript{Mo}$ is about half of that the semiconducting \emph{2H} phase. 
We suggest that the reason for this is twofold. 
On the one hand, similarly to S vacancies, atoms can be more easily removed from the lattice due to the different formation energy of the pristine single-layers. On the other hand, removing a Mo atom from \emph{1T'} lattice induces a strong reconstruction (see Fig. \ref{VMoS}e), in which one S atom moves towards the vacant site and forms covalent bonds with two S atoms in the opposite chalcogen layers. This bond is 2.13 {\AA} long, very close to the value observed in the puckered S$_8$ rings, as mentioned in Section II. The formation of this bond compensates for the dangling bonds otherwise formed, \emph{i.e.} the main origin of the high formation energy of the \emph{2H}-V$\textsubscript{Mo}$ defect. In this latter phase, such a structural relaxation is not observed (Fig. \ref{VMoS}d), and the removal of the metal atom leaves the lattice almost unperturbed.

To gain further insight into the role of S--S bonds forming upon defect reconstruction, we consider a composite \emph{1T'}-V$\textsubscript{MoS}$ defect in which the central S atom is removed. Its atomic structure is shown in Fig.~\ref{VMoS}f, and presents a new bond between S atoms belonging to opposite sulfur atomic planes. It is worth noting that formation energy of \emph{1T'}-V$\textsubscript{MoS}$ defect (Fig.~\ref{FE-Vacancies}) is lower than the sum of formation energies of single Mo and S vacancies by 1.14~eV. This suggests that it is more likely to form a S vacancy within the coordination network of a Mo vacancy  than in the clean area of the monolayer.

Figs.~\ref{DOS-Vacancies}c,d show the density of states of MoS$_2$ with Mo vacancy defects. In the \emph{2H} phase, \emph{2H}-V$\textsubscript{Mo}$ leads to the formation of three deep levels, two of them due to the sulfur orbitals and one composed of Mo orbitals, whereas in \emph{1T'}-MoS$_2$ one can recognize defect-induced states approximately 0.1~eV and 0.5~eV below the Fermi level.

In general, from Fig. \ref{FE-Vacancies}, one can notice a sharp contrast between the stability of single vacancies in the two crystalline phases. In the \emph{2H} phase, the sulfur vacancy is the most stable defect for the entire range of  chemical potential $\mu_S$. Conversely, in the \emph{1T'} phase, under extreme S-rich conditions (\emph{i.e.} $\mu_S$ = $-$1.25~eV), \emph{1T'}-V$\textsubscript{Mo}$ is more stable than \emph{1T'}-V$\textsubscript{S1}$ by 0.15~eV. Overall, from the smaller formation energy for \emph{1T'}-V$\textsubscript{Mo}$ defect in the \emph{1T'} phase compared to the \emph{2H} phase one can expect that Mo vacancies in the semimetallic phase are more abundant than  in the semiconducting phase.
%  In the \emph{2H} phase, sulfur-based multiple defects form because of the  stability of  \emph{2H}-V$\textsubscript{S}$, whereas the unlikely \emph{2H}-V$\textsubscript{Mo}$ formation makes Mo deficient multiple vacancies very unlikely to occur. The situation could be different for \emph{1T'}-MoS$_2$, where a much larger relative stability of  \emph{1T'}-V$\textsubscript{Mo}$ defect is observed.

We would like to stress that in our work all calculations are performed on charge-neutral states of defects. Previous theoretical works addressed the charged defects in semiconducting $2H$-MoS$_2$ and found that both $2H$-V\textsubscript{S} and $2H$-V\textsubscript{Mo} are likely to assume negative charge states when the Fermi level is close to the conduction band (\emph{i.e.} $n$-type doping)\cite{kom15,noh14}. However, because of the semimetallic character of $1T'$-MoS$_2$, defects in this material have been considered only in their neutral state and, for consistency of comparing the two phases, equivalent defects in the $2H$-MoS$_2$ are also considered neutral. For a detailed discussion of charge transition levels in semiconducting MoS$_2$ the reader is referred to Refs.~\onlinecite{kom15,noh14}.

%ADATOMS%
\subsection{Adatom Defects}

\begin{figure}[]
  \centering
  \includegraphics[width=1\columnwidth]{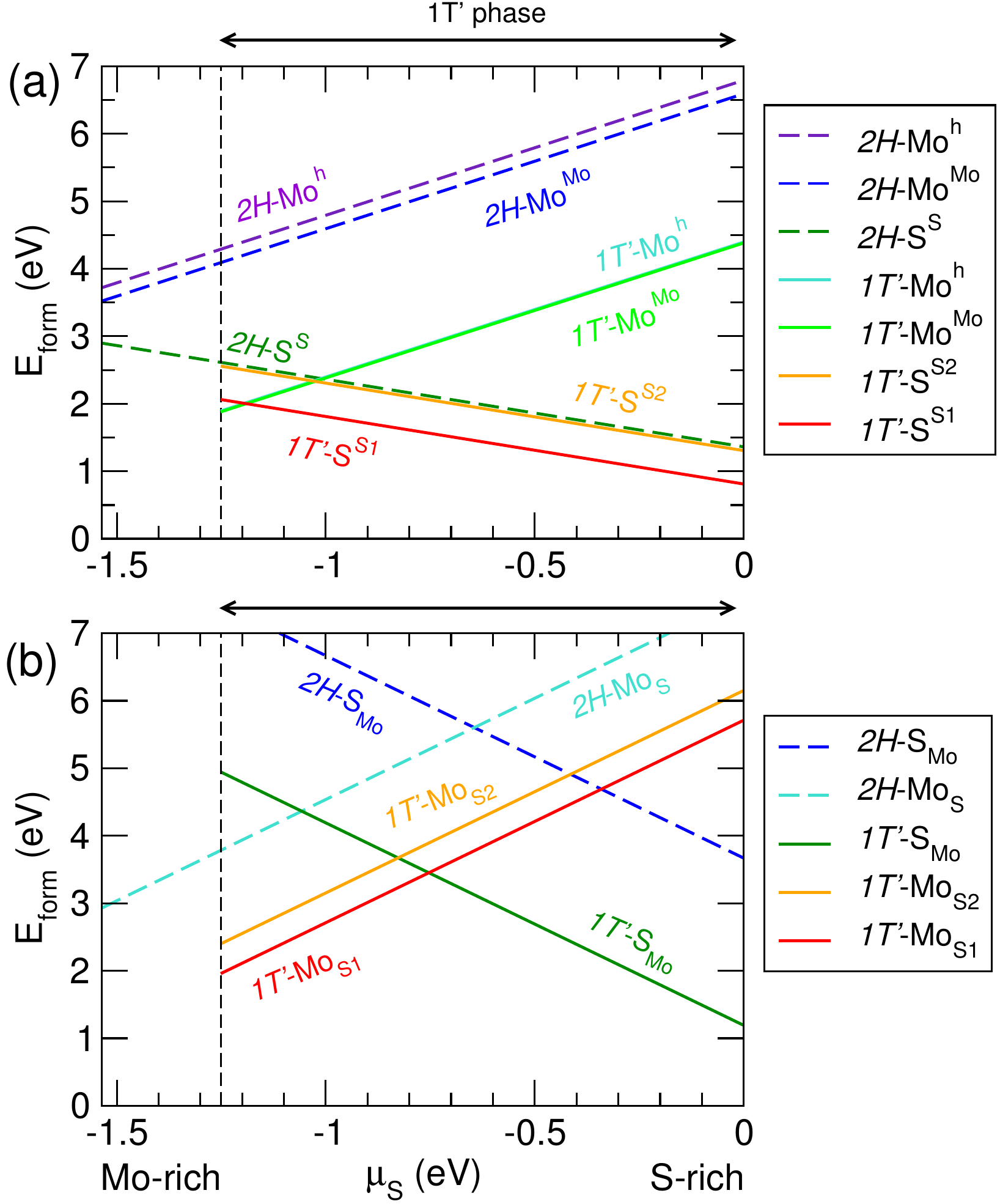}
  \caption{Formation energies of (a) adatom and (b) antisite defects as a function of chemical potential of sulfur $\mu_S$. Vertical dashed line indicates the range of stability of \emph{1T'}-MoS$_{2}$. Colored solid (dashed) lines represent the formation energies of defects in the \emph{1T'} (\emph{2H})  phase. \label{FE-AdatomsAntisites}}
\end{figure}

Next, we examine S  and Mo adatoms  on MoS$_2$ monolayers  and the impact these defects have on the electronic and magnetic properties of the two phases. Chemisorption of a Mo (S) atom is thermodynamically more favorable in a Mo-rich (S-rich) conditions, as seen in Fig. \ref{FE-AdatomsAntisites}a.

For S adatoms in the \emph{2H} phase, we considered three different adsorption sites, namely on top of a Mo atom, on top of a S atom and on the hollow site. The S adatom is stable only when bound on top of a S atom (\emph{2H}-S$\textsuperscript{S}$), forming a S--S bond of 1.96~{\AA}, in excellent agreement with previous reports\cite{kom15, cao16}. Such an interatomic distance, very close to the one of 1.94~{\AA} calculated for  the S$_2$ molecule, clearly signals the formation of a strong covalent bond. This is corroborated by the low formation energy,  among the lowest for all  impurities considered in this work  for the \emph{2H} phase\cite{noh14}. 
In the case of \emph{1T'} polymorph, S adatoms on top of both S1 (\emph{1T'}-S$\textsuperscript{S1}$) and S2 (\emph{1T'}-S$\textsuperscript{S2}$) sites were considered. The resulting S--S bond lengths are 1.98~{\AA} and 1.95~{\AA} for \emph{1T'}-S$\textsuperscript{S1}$ and \emph{1T'}-S$\textsuperscript{S2}$, respectively, very similar to the ones observed for adsorption on the \emph{2H} phase. The defect formation energy for S adatom on top of S1 is 0.49 eV~lower than on the S2 site.  
In general, one can say that the S1 is not only the most prone lattice site to forming vacancies, but it is also the most reactive towards chemisorption, whereas the S2 site in \emph{1T'} phase behaves more similarly to the S atom in the \emph{2H} in terms of defects energetics, both with respect to sulfur vacancy as well as adatom formation.

\begin{figure}[]
  \centering
  \includegraphics[width=1\columnwidth]{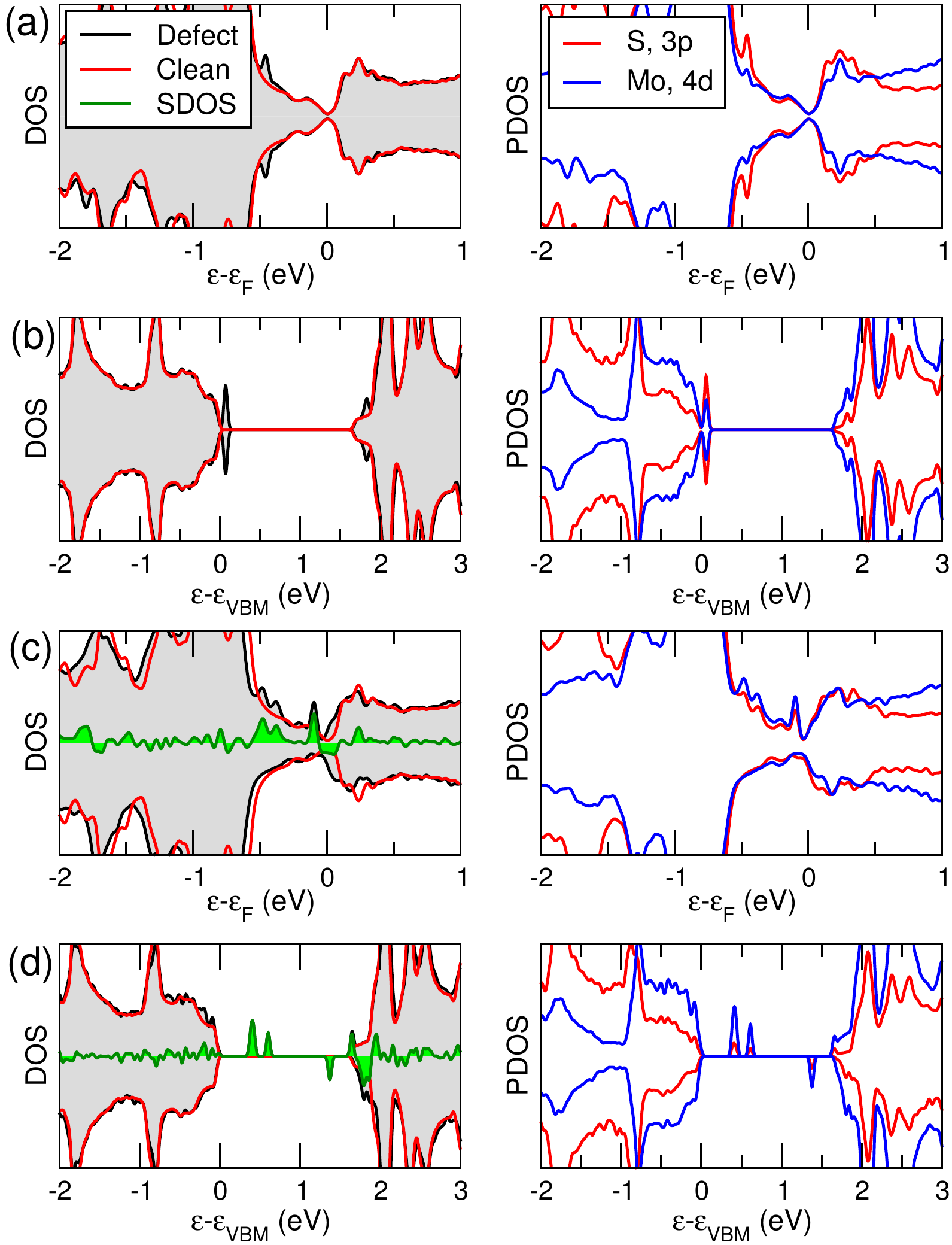}
  \caption{Total (left) and projected (right) electronic density of states plots of (a) \emph{1T'}-S$\textsuperscript{S1}$, (b) \emph{2H}-S$\textsuperscript{S}$, (c) \emph{1T'}-Mo$\textsuperscript{Mo}$  and (d) \emph{2H}-Mo$\textsuperscript{Mo}$ defects. The total DOS of the pristine (red lines) and the defective (black lines) monolayers  are superimposed. Green shaded area indicates the net spin-polarization (SDOS), defined as the difference between spin-majority and spin-minority DOS.  \label{DOS-Adatoms}}
\end{figure}

Despites the very different thermodynamic stability and electronic structure of the monolayers as well as the energetics involved in the S adsorption presented above, comparable changes in the density of states upon the formation of this impurity are found. Specifically,  from Fig. \ref{DOS-Adatoms}a,b, for both  phases one can observe that the adatom-induced states are only due to the sulfur $3p$ atomic orbitals. For the \emph{2H} phase, two localized states appear in the band gap, one very close to the valence band edge and the other close to the conduction band edge, indicating that S adatom acts as a shallow defect. In the \emph{1T'} phase, the chalcogen chemisorption leads to a sharp peak located at {\it ca.} 0.5 eV below the Fermi level.

We then investigated configurations of Mo adatom on top of a Mo site (Mo$\textsuperscript{Mo}$) as well as on the hollow site (Mo$\textsuperscript{h})$. In the case of \emph{2H} phase, these configurations differ by 0.20~eV and in general represent the most energetically unfavorable defects among the ones presented in Fig. \ref{FE-AdatomsAntisites}a\cite{kom13}. In the \emph{2H}-Mo$\textsuperscript{Mo}$ configuration, the adatom is located 3.06~{\AA} above the Mo layer, with an average distance with the three neighboring S atoms equals to $2.46$ {\AA}, close to the interatomic distances observed in the \emph{2H}-Mo$\textsuperscript{h}$ configuration (2.45 {\AA})\cite{ata11bis}.
Mo adsorption  on the \emph{1T'} phase is considerably more energetically favorable compared to the \emph{2H} phase by $\approx$2~eV, therefore suggesting that the formation of Mo species on such monolayer are more likely to occur in the former polymorph compared to the latter. Additionally, adsorption on top of Mo site is only slightly more stable than on hollow site (by 13~meV), although the corresponding bond lengths are overall similar to the one observed for Mo adsorbed on \emph{2H}-MoS$_2$.

Contrary to the \emph{1T'} phase, where S adatom is the most stable defect for the entire  range of stability of the monolayer, in the \emph{1T'} phase extreme Mo-rich conditions favors the formation of Mo adatoms, both on top of a Mo atom as well as on the hollow site (by 0.18 eV and 0.17 eV, respectively), as it can be noticed at $\mu_S$ = $-$1.25 eV in Fig. \ref{FE-AdatomsAntisites}a.

In Figs. \ref{DOS-Adatoms}c,d we show the density of states of Mo adatom defects. In the \emph{2H} phase, the adatom induces three states in the band gap: one close to the conduction band edge and two located at $\approx$0.5 eV above the valence band maximum. This is in contrast to what is observed for Mo adatoms on \emph{1T'}-MoS$_2$, where Mo adatoms give rise to a localized states at the Fermi energy. 

\begin{figure}[]
  \centering
  \includegraphics[width=1\columnwidth]{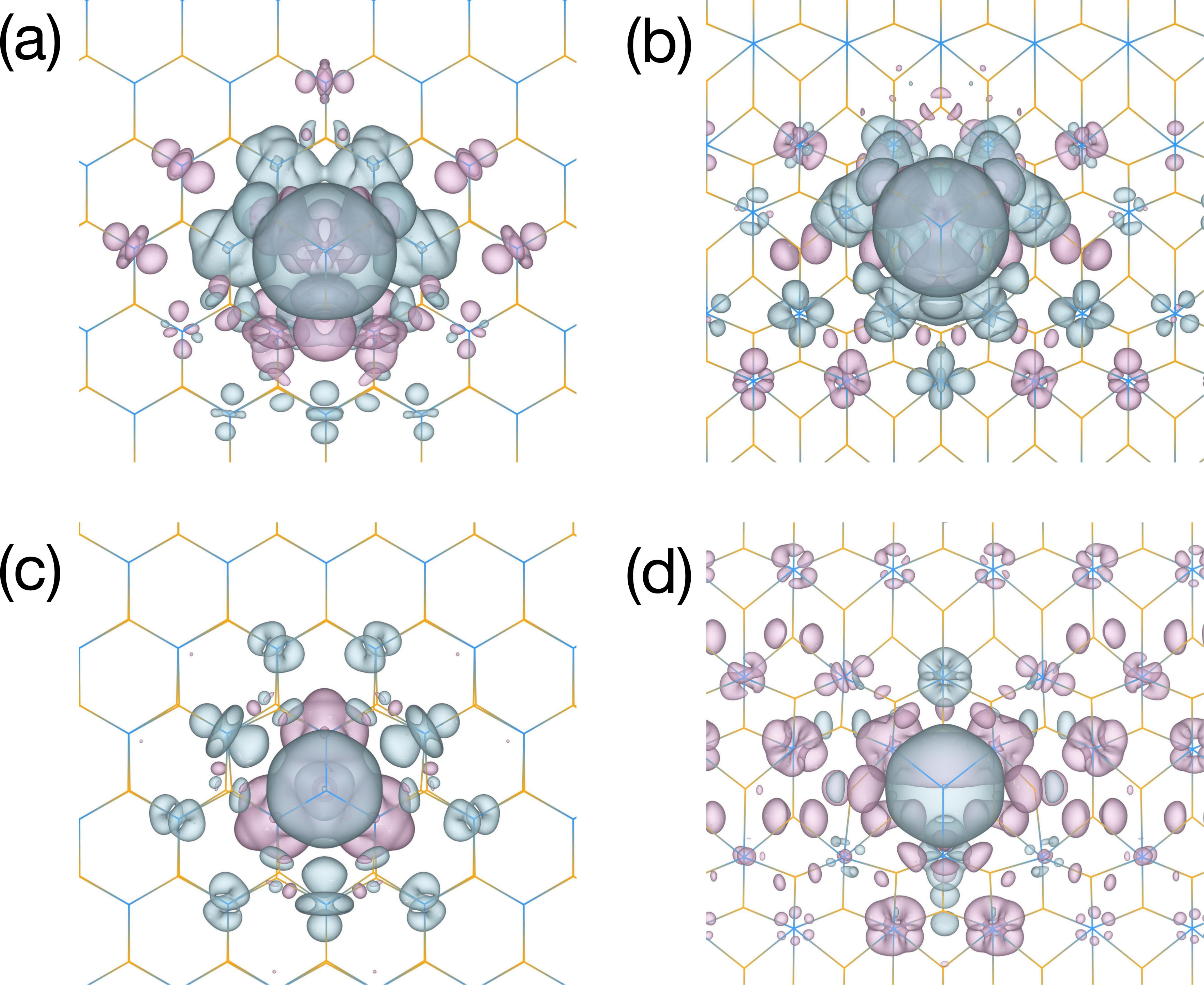}
  \caption{Spin-densities upon (a)\emph{2H}-Mo$\textsuperscript{Mo}$, (b) \emph{1T'}-Mo$\textsuperscript{Mo}$, (c) \emph{2H}-Mo$\textsubscript{S}$ and (d) \emph{1T'}-Mo$\textsubscript{S}$ defects formation. Grey (pink) clouds represent spin-majority (spin-minority) isosurfaces, with isocontours  of $\pm$ 0.003 (0.008) \emph{e} \AA$^{-3}$ for adatoms (antisites). \label{SpinDensities}}
\end{figure}

Despite this marked difference, Mo chemisorption leads to a large magnetic moment $\mu$ = 4 $\mu_{B}$ in the lattice, irrespective of the crystalline phase and defect configuration\cite{cao16, kom15}. We address the details of defect-induced magnetism in Fig. \ref{SpinDensities}a,b. This figure shows the spin densities at Mo adatoms on the top site (Mo$\textsuperscript{Mo}$) for the two polymorphs.
%, \emph{i.e.} the difference in electron density for spin-majority and spin-minority contributions. 
In the \emph{2H} phase, the spin-density is strongly localized around the defect, whereas in \emph{1T'}  appears to be spread out  due to the hybridization of the spin-polarized localized states with the delocalized electrons of the semimetallic phase. This can be observed also from the green shaded area of Figs. \ref{DOS-Adatoms}c,d, where, contrary to \emph{1T'}-MoS$_2$, in the semiconducting phase magnetism is mainly due to the mid-gap states. Because of the relative low formation energy necessary to host a Mo adatom in the \emph{1T'} polymorph and its stability in Mo-rich conditions, Mo adsorption may offers a  feasible way towards magnetism in two-dimensional MoS$_{2}$ with potential application in spintronics.

We further calculate the magnetocrystalline anisotropy energies $E_{\rm MAE}$ of the stable configurations of Mo adatom defects, that turn out to be $-$0.85~meV and 0.11~meV for $2H$-Mo\textsuperscript{Mo} and $1T'$-Mo\textsuperscript{Mo} defects, respectively. This implies that the magnetic moment easy axis is in-plane (out-of-plane) for adatom defects in single-layer $2H$- ($1T'$-) MoS$_2$.

%ANTISITES%

\subsection{Antisite Defects}
\begin{figure}[b]
  \centering
  \includegraphics[width=0.65\columnwidth]{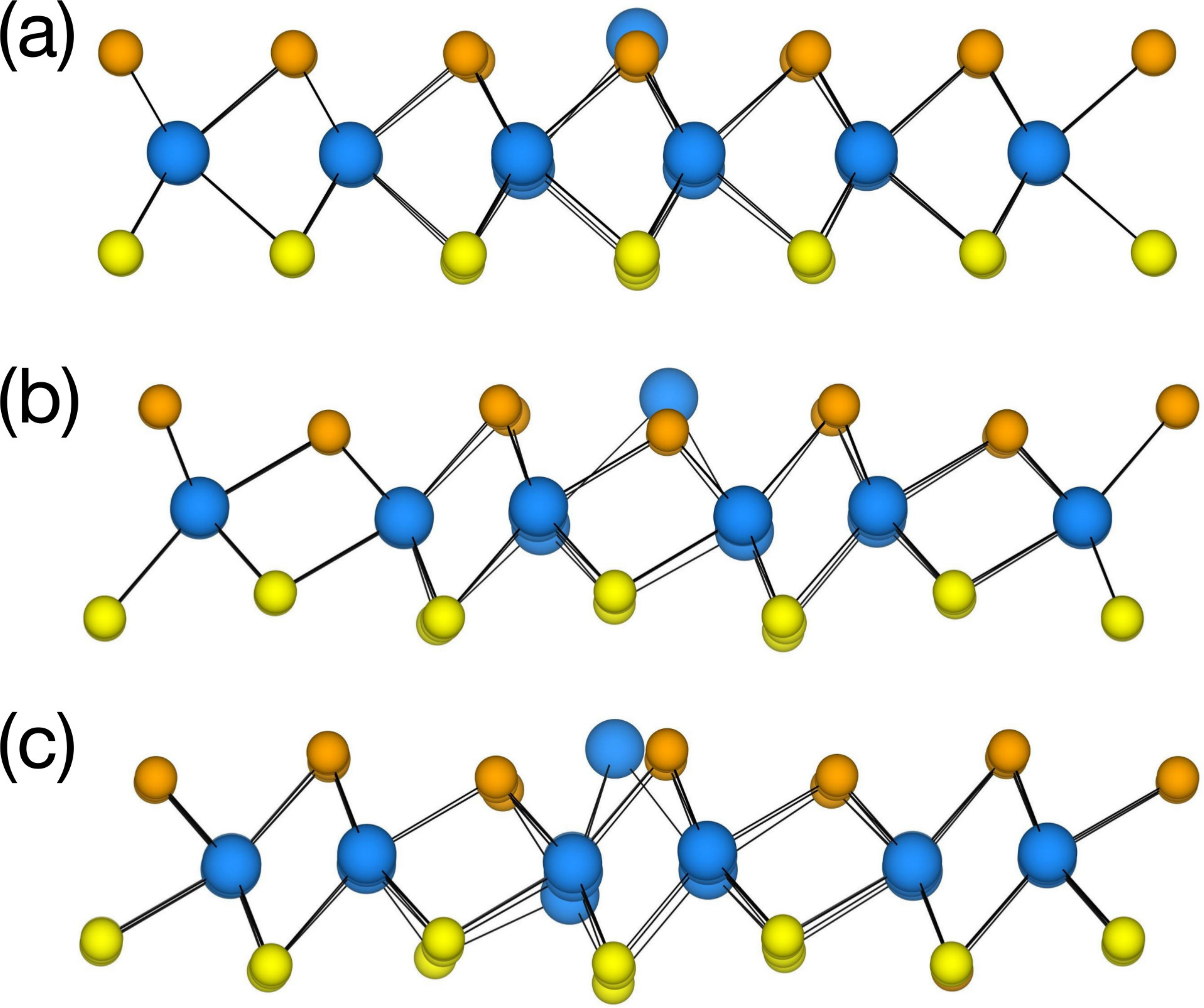}
  \caption{Atomic structures of  (a) \emph{2H}-Mo$\textsubscript{S}$, (b) \emph{1T'}-Mo$\textsubscript{S1}$ and (c) \emph{1T'}-Mo$\textsubscript{S2}$ defects.\label{Antisites} }
\end{figure}

As a final subject of investigation, we discuss antisite defects. These defects form upon the replacement of a Mo (S) atom with a  S (Mo) atom. Therefore, in order to rationalize our results, we decompose the antisite defect formation into a  hypothetical two-steps process, namely \emph{(i)} the vacancy formation and \emph{(ii)} the addition of an adatom of the element complementary to the removed atom.  Bearing this idea in mind, we can understand our results in the light of the formation energies presented in the previous two sections. One can expect that the more likely the vacancy and adatom formation, the more likely the antisite defect.
As pointed out in Fig. \ref{FE-AdatomsAntisites}b,  Mo-rich conditions stabilize both S vacancies and Mo adatoms, and therefore would favor the formation of Mo$\textsubscript{S}$ antisites. Reciprocally, S-rich conditions favour both the formation of Mo vacancies and S adatoms, hence S$\textsubscript{Mo}$ antisite defects would be stabilized.

In both phases, upon the S$\textsubscript{Mo}$ formation, the substituting S atom is surrounded by other six sulfur atoms, because of the trigonal pristmatic coordination of the Mo atom in the pristine lattice. In the \emph{2H} phase, the bond distance of the impurity with the neighboring atoms is 2.40 {\AA}, slightly shorter than the Mo--S bond length observed in the pristine crystal. Similarly, in \emph{1T'}-MoS$_2$, the same average bond length is 2.32 {\AA}, even shorter than typical Mo--S distances in this polymorph. Such a bond contraction at the impurity site can be interpreted in term of covalent radii: since S atom is smaller (1.05~{\AA}) than Mo (1.54~{\AA}), the atoms surrounding the imperfection move towards the defective site in order to accommodate the chalcogen impurity, thus locally shortening the bond distances.
S$\textsubscript{Mo}$ antisites show a very different formation energy in the two crystalline phases, being about 2~eV more favorable in the \emph{1T'} phase compared to the \emph{2H} phase at the same value of chemical potential $\mu_S$. This is not surprising since both Mo vacancy formation as well as S adsorption are more favorable in the \emph{1T'} phase compared to the \emph{2H} phase. Comparing the formation energy of S$\textsubscript{Mo}$ antisites with the sum of the formation energies of isolated Mo vacancy and S adatom, it is found that such antisite is more stable by 2.35~eV in the \emph{2H} phase and 1.26~eV in the  \emph{1T'} phase, thereby indicating a remarkable reactivity of the Mo vacant site.

\begin{figure}[]
  \centering
  \includegraphics[width=1\columnwidth]{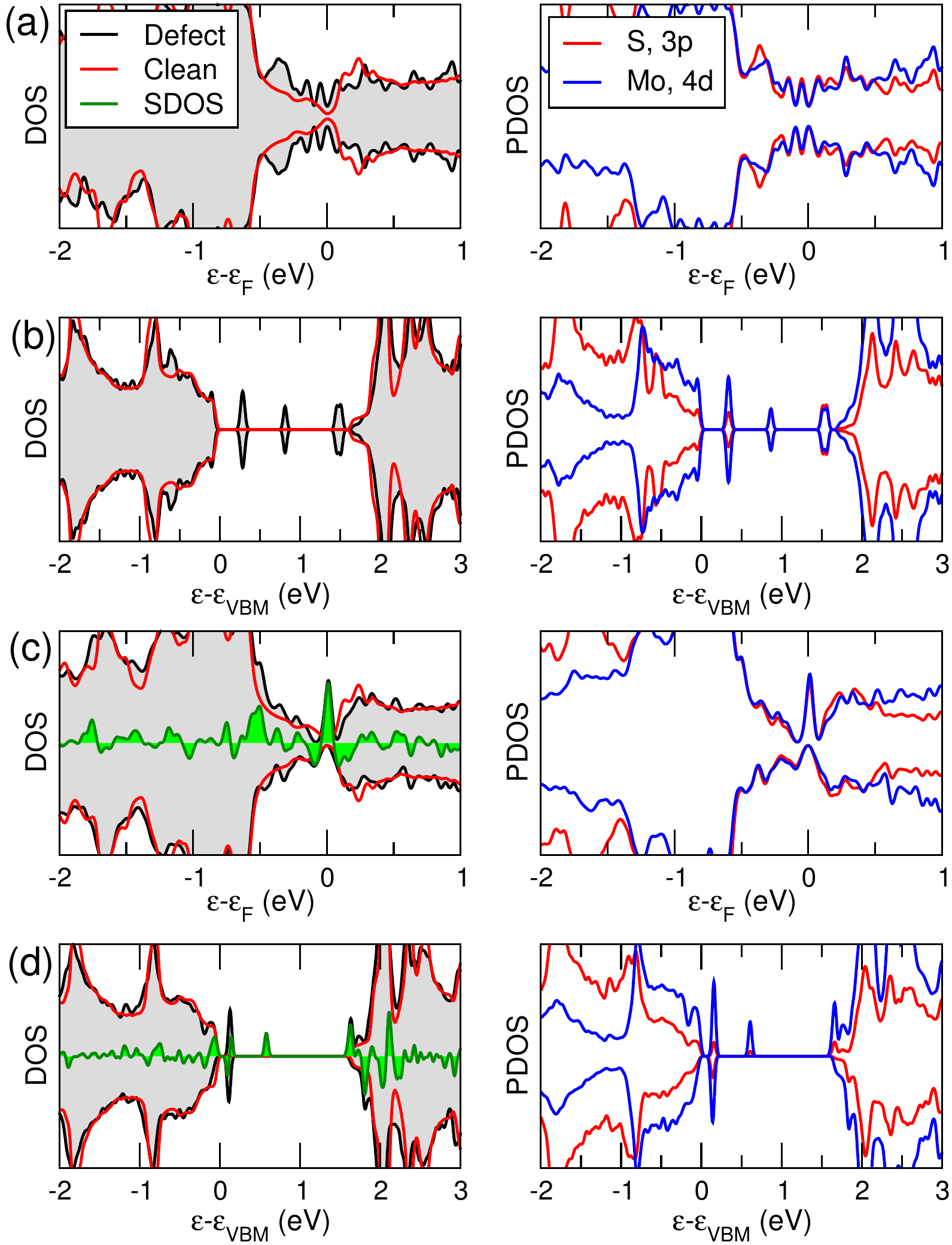}
  \caption{ Total (left) and projected (right) electronic density of states plots of (a) \emph{1T'}-S1$\textsubscript{Mo}$, (b) \emph{2H}-S$\textsubscript{Mo}$, (c) \emph{1T'}-Mo$\textsubscript{S1}$  and (d) \emph{2H}-Mo$\textsubscript{S}$ defects. The total DOS of the pristine (red lines) and the defective (black lines) monolayers  are superimposed. Green shaded area indicates the net spin-polarization (SDOS), defined as the difference between spin-majority and spin-minority DOS. \label{DOS-Antisites} }
\end{figure}

The S$\textsubscript{Mo}$ antisites manifest in the electronic structure (Fig. \ref{DOS-Antisites}a,b) as three defect-induced states. In the case of \emph{2H} phase, two of these peaks are located close to the band edges, whereas the third one is in the middle of the band-gap. In \emph{1T'} phase, the three impurity states are occupied and form around the Fermi level. Similarly to \emph{2H}-MoS$_2$, their composition  is mainly due to the $4d$ Mo states, as a consequence of the change in the Mo--Mo interaction that occurs when the substitution of the metal with the chalcogen atom takes place. As we already observed for Mo-vacancies and S-adatoms described above, S$\textsubscript{Mo}$ defect does not lead to any spin-polarization\cite{cao16}.

Finally, we study the formation of Mo$\textsubscript{S}$ antisites shown in (Fig. \ref{FE-AdatomsAntisites}b). 
In the \emph{2H} phase, there is only one possible configuration for this defect,  whereas two distinct structures can form in the \emph{1T'} polymorph, depending on whether substitution takes place in position S1 (\emph{1T'}-Mo$\textsubscript{S1}$) or position S2 (\emph{1T'}-Mo$\textsubscript{S2}$). The atomic structures presented in Fig. \ref{Antisites} show that the substitution of a S atom for  Mo atom leads to a distortion in the lattice due to the larger atomic radius of Mo atom compared to the S atom.  The distortion is more pronounced for the \emph{1T'} phase, in particular in the \emph{1T'}-Mo$\textsubscript{S2}$ configuration shown in Fig.  \ref{Antisites}c where the larger metal atom can barely be accommodated between two Mo atoms in the metallic layer, whose Mo--Mo average distance in the pristine system is only 2.92 {\AA}. In the \emph{2H} phase, the resulting bond lengths between the substitutional Mo at the defective site and the three surroundings Mo are equal 2.75 {\AA}. In the \emph{1T'} phase, the same three bond lengths are 3.05 {\AA} for \emph{1T'}-Mo$\textsubscript{S1}$, whereas in \emph{1T'}-Mo$\textsubscript{S2}$ two of these bond lengths are equal to  3.01 {\AA} and the other to 2.56 {\AA}.
The formation energies of these antisites are shown in Fig. \ref{FE-AdatomsAntisites}b. Similarly to all the  defects discussed above, Mo$\textsubscript{S}$ antisites  preferentially form in the  \emph{1T'} phase than in the \emph{2H}, and, again, defects involving S1 sites (\emph{i.e.} \emph{1T'}-Mo$\textsubscript{S1}$) are more stable than the one involving S2 (\emph{i.e.} \emph{1T'}-Mo$\textsubscript{S2}$), with an energy difference of 0.44 eV. This is consistent with our initial assumption based on the fact that both vacancies and adatoms have lower formation energies in position S1 rather than in position S2. In analogy with the S$\textsubscript{Mo}$ defect, \emph{2H}-Mo$\textsubscript{S}$ antisite is more stable than isolated Mo adatom and S vacancy by 2.08~eV, while in \emph{1T'} phase this value turns out to be 1.31~eV and 2.11~eV for \emph{1T'}-Mo$\textsubscript{S1}$ and \emph{1T'}-Mo$\textsubscript{S2}$, respectively. 

%%Overall, the stability of the antisite defects depends strongly on the chemical potential. As shown in Fig. \ref{FE-AdatomsAntisites}b, for both phases, Mo$\textsubscript{S}$ antisites are stable in Mo-rich conditions, whereas S- rich environment leads to the enhancement of the stability of S$\textsubscript{Mo}$ defects.

The effect of Mo$\textsubscript{S}$ antisites on the density of states is shown in Figs. \ref{DOS-Antisites}c,d.  Similarly to the S$\textsubscript{Mo}$ defect previously discussed, the formation of this antisite defect in the semiconducting phase leads to the formation of three mid-gap states, two closer to the band edges and one deep the gap. In the \emph{1T'}  phase, though, the main consequence of such impurity is a sharp level located around the Fermi level, similarly to what was observed for Mo adatoms.

A further similarity with Mo adatom is the magnetism induced by Mo$\textsubscript{S}$ antisite in both crystalline phases\cite{cao16}. Contrary to adatoms, however, the resulting magnetic moment is $\mu$ = 2 $\mu\textsubscript{B}$. The net spin-polarization is shown as green shaded area in Figs. \ref{DOS-Antisites}c,d, where one can notice that in both phases the origin of magnetism  is mainly due to defect states, but in the \emph{1T'}  such impurity levels appear hybridized with the bulk electronic states of the monolayer. This can be also observed in Figs. \ref{SpinDensities}c,d, where the spin-density is more spatially confined around the imperfection in the semiconducting phase compared to the semimetallic phase, where some spin-density is accommodated also at the neighboring sites.

In addition, we find $E_{\rm MAE}$ equal to $-$1.31~meV and 0.53~meV for antisite defects in $2H$-Mo\textsubscript{S} and $1T'$-Mo\textsubscript{S1}, respectively, thereby suggesting that, similarly to what we observed for Mo adatoms, the easy axis is located in the in-plane (out-of-plane) direction for such antisite defects in the $2H$ ($1T'$) phase.

% CONCLUSION

\section{Conclusion}
We performed a comprehensive first-principles investigation of the atomic structure and thermodynamic stability of intrinsic point defects in the semimetallic \emph{1T'}  phase of single-layer MoS$_2$ and made a thorough comparison with equivalent defects in the semiconducting \emph{2H} phase. We explored a large number of defects among vacancies, adatoms as well as antisites and examined their impact on the electronic and magnetic properties in each of the crystalline phases.

Our simulations clearly indicate that all considered defects present lower formation energies in the metastable \emph{1T'} phase compared to the \emph{2H}. Therefore, under thermodynamic equilibrium, \emph{1T'} polymorph
is expected to be more susceptible to lattice imperfections. The reason for this can be traced back to the lower stability of the \emph{1T'} monolayer with respect to \emph{2H}-MoS$_2$. Specifically, our findings also suggest that impurities preferentially form at the S atom closer to the Mo atomic plane.  Comparing the formation energy of all investigated point defects in the \emph{1T'} phase, we conclude that  similarly to the \emph{2H} phase the most stable impurities are sulfur vacancy defect (\emph{1T'}-V$\textsubscript{S1}$) in Mo-rich conditions and sulfur adatoms in S-rich conditions (\emph{1T'}-S$\textsuperscript{S1}$).

Among vacancy and adatom defects, in \emph{1T'}-MoS$_2$ extreme Mo-rich conditions promote the formation of \emph{1T'}-V$\textsubscript{Mo}$ and \emph{1T'}-Mo$\textsuperscript{Mo}$, in sharp contrast to \emph{2H}-MoS$_2$, where  \emph{2H}-V$\textsubscript{S}$ and  \emph{2H}-S$\textsuperscript{S}$ remain stable in the whole range of chemical potential. Concerning antisite defects, a qualitatively similar behavior is observed in the two phases, in a sense that S-rich (Mo-rich) conditions stabilize S\textsubscript{Mo} (Mo\textsubscript{S}) defects, though such impurities present lower formation energies in the \emph{1T}' phase compared to the \emph{2H}.

Irrespective of the crystalline phase, our simulations further indicate that Mo adatoms and Mo antisites lead to the formation of a local magnetic moment, thereby suggesting that such defects are likely to be responsible for the magnetism experimentally observed upon proton irradiation of MoS$_2$ flakes\cite{mat12}. 
Furthermore, such defect-induced magnetic moments have in-plane (out-of-plane) easy axis when form in the $2H$ ($1T'$) phase.
On the other hand, S and Mo vacancies as well as S adatoms and antisites do not results in any spin-polarization. Due to the semimetallic nature of the \emph{1T'} phase, the defect-induced electronic levels are partially hybridized with the bulk states whereas they show up as sharp mid-gap states in the \emph{2H} phase.

Overall, our qualitative conclusions can be extended to the other group-VI single-layer disulfides and diselenides such as MoSe$_2$, WS$_2$ and  WSe$_2$ where similar relative stabilities of the two structural phases and electronic properties were observed.

% ACKNOWLDEGMENTS
 
\begin{acknowledgments}
M. P. gratefully acknowledges Fernando Gargiulo and Davide Colleoni for fruitful discussions. This work was supported by the Swiss National Science Foundation (grant No. 200021\_162612). All first-principles calculations were performed at the Swiss National Supercomputing Centre (CSCS) under the project s675.
\end{acknowledgments}

%\bibliography{References}

\end{document}